\documentclass[aps,showpacs,twocolumn,superscriptaddress]{revtex4}
\usepackage{graphicx}
\begin{document}
\title{Reflection symmetry instability at high spins in $^{162,164}$Yb}
\author{R. G.~Nazmitdinov}
\affiliation{Departament de F{\'\i}sica, 
Universitat de les Illes Balears, E-07122 Palma de Mallorca, Spain}
\affiliation{Bogoliubov Laboratory of Theoretical Physics,
Joint Institute for Nuclear Research, 141980 Dubna, Russia}
\author{J. Kvasil}
\affiliation{Institute of Particle and Nuclear Physics, Charles
University, V.Hole\v sovi\v ck\'ach 2, CZ-18000 Praha 8, Czech Republic}
\author{A. Tsvetkov}
\affiliation{Institute of Particle and Nuclear Physics, Charles
University, V.Hole\v sovi\v ck\'ach 2, CZ-18000 Praha 8, Czech Republic}
\date{\today}
\begin{abstract}
A shape evolution of $^{162,164}$Yb in yrast states is traced
using  the self-consistent Skyrme Hartree-Fock calculations.  
We found that  nonaxial octupole deformations (in particular, $Y_{31}$ term)  
become favorable at $\hbar \Omega > 0.4$MeV in  $^{162}$Yb, while in  $^{164}$Yb 
a nonaxial quadrupole shape is dominant at fast rotation.
The cranked Nilsson model and random phase approximation are used to understand
the dynamics of octupole correlations in both nuclei. We demonstrate that 
the disappearance of one of the octupole vibrational modes in the rotating frame gives rise to
the nonaxial octupole deformations in $^{162}$Yb, while the octupole modes are nonzero in
 $^{164}$Yb.
\end{abstract}
\maketitle

Shell structure plays a prominent role in the formation of ground  and excited nuclear 
states. It  is crucially important at fast rotation, when nuclear stability against 
fission is determined by a delicate balance of different constituents of a nuclear 
potential. In a geometrical approach \cite{BM75},  where the concept of shape 
deformation is one of the basic cornerstones, effects produced by quadrupole degrees 
of freedom are well understood for various effective potentials. 
Whether  octupole degrees of freedom are of any importance and how they are
manifested,-- these questions stimulate a noticeable fraction of experimental efforts
in high spin physics \cite{lit}. 

Numerous calculations exploiting the intrinsic reflection 
symmetry breaking of a nuclear mean field predict the existence 
of stable axial octupole deformation in the ground state 
for Ra-Th $(Z\sim 88$, $N\sim 134$) and Ba-Sm $(Z\sim 58$, $N\sim88$) nuclei, where
strong octupole instability is  due to the coupling of 
$j_{15/2} - g_{9/2}$, $i_{13/2} - f_{7/2}$, $h_{11/2} - d_{5/2}$ orbitals \cite{Bu96}. 
The potential energy surface produced by different calculations is, however, quite shallow. 
Experimental nuclear spectra also show a pronounced parity splitting at low angular 
momenta. Rather dynamical octupole effects (vibrations) are dominant over static 
effects (octupole deformations) at low spins. The increase of the angular momentum 
decreases pairing correlations which reduce the octupole interaction, since they 
couple the orbitals with the same parity. With rotation, the density of different 
parity states is enhancing markedly near the Fermi level. Indeed, one observes a 
smooth decrease of the parity splitting with the increase of the rotational frequency 
fairly in a few nuclei. These features are nicely reproduced  within the 
cranking+Hartree-Fock-Bogoliubov (HFB) approach with Gogny forces in Ba-Sm 
region \cite{Ga98}. Similar calculations with Skyrme \cite{Ya01} and Gogny \cite{Ta01} 
forces predict a nonaxial $Y_{31}$ octupole deformation in light nuclei at high spins.
The results based on the Skyrme interaction demonstrate the importance of a nonaxial 
$Y_{32}$ octupole deformation in actinide nuclei at fast rotation  \cite{Ts02}. 
Notice that rotation induces the contribution of nonaxial quadrupole and octupole 
components of effective nuclear potentials.

A wealth of experimental information accumulated in the past few years  on high 
spin states in the transitional nuclei $Z\sim 70$, $N\sim 90$ prompted us to 
investigate the question to what extent dynamical octupole effects observed at 
low spins in this mass region might develop static octupole deformations at 
sufficiently high angular momenta. This is  a main subject of  this Letter.

To analyse the rotational evolution of equilibrium deformations we have employed 
the HFODD code \cite{Do05}. In the present calculations the pairing correlations 
are neglected, since they are not expected to play an important role at large 
rotational frequencies. Our HF calculations based on the cranked Skyrme SKP 
interaction (see details in Ref.\onlinecite{Ts02}) demonstrate that a spontaneous 
symmetry breaking phenomenon occurs in $^{162}$Yb at large rotational 
frequencies (see Fig.1). 
\begin{figure}[ht]
\includegraphics[height=0.17\textheight,clip]{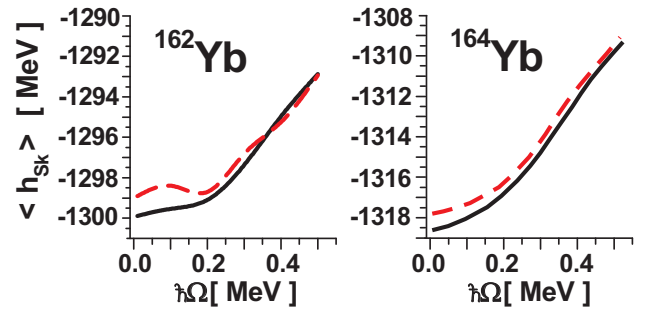}
\caption{(Color online) The rotational evolution of the SkP Skyrme mean field energy: 
pure nonaxial quadrupole (solid line)
and  nonaxial quadrupole+octupole (dashed line)  solutions. 
} 
\label{fig1}
\end{figure}
\begin{figure}[ht]
\includegraphics[height=0.42\textheight,clip]{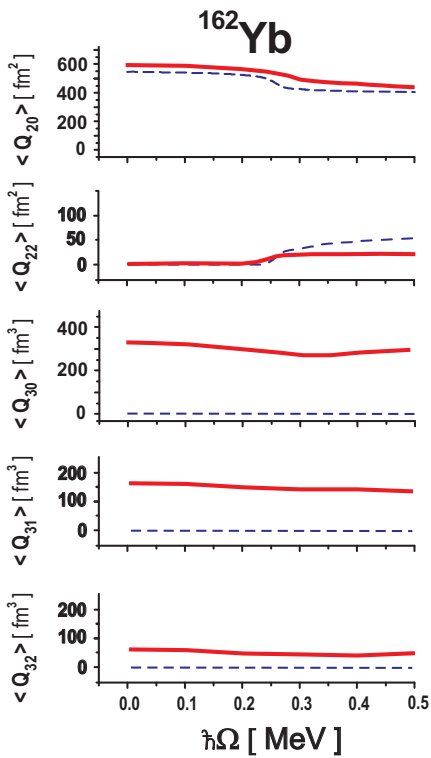}
\caption{(Color online) The rotational behaviour of  quadrupole $\langle Q_{2m}\rangle$ 
and octupole $\langle Q_{3m}\rangle$ momenta for a broken 
(solid line) and an unbroken (dashed line) reflection symmetry  SkP Skyrme mean 
field solutions for $^{162}$Yb.
} 
\label{fig2}
\end{figure}
The  mean field solution with a broken reflection symmetry 
(with nonzero $\langle Q_{30}\rangle$, $\langle Q_{31}\rangle$  momenta and
a small admixture of $\langle Q_{32}\rangle$ one, see Fig.2) 
becomes favorable at $\hbar \Omega > 0.4$MeV, in contrast to the solution with 
a reflection symmetry. In $^{164}$Yb  both the minimal solutions are 
very close to each other, while the pure quadrupole one determines properties of 
yrast states at fast rotation. Hereafter, $\langle...\rangle$ means the averaging 
over the mean field vacuum (yrast) state at a given rotational frequency $\Omega$. 
We obtain a similar pattern with the SKIII parametrisation, which conforms our main 
finding, however, with slightly higher values of the energy minima. 
At slow 
rotation $\hbar\Omega\leq 0.3$MeV (when pairing is important) the results should 
be taken cautiously, while  the calculations provide a most credible tendency 
at $\hbar\Omega>0.3$MeV.

A transparent physical idea that an instability of a nuclear potential  with
respect to  a  given deformation implies a softening of the corresponding 
vibrational mode \cite{BM75} enables us to shed light on the mean field results.
The cranked shell model that incorporates 
a random phase approximation (RPA)  represents a powerful tool to illuminate 
dynamics of various correlations, evolving from shape 
fluctuations (vibrations) to a nuclear shape instability.
We use a CRPA approach \cite{Kv06} that consists in 
a self-consistent solution of 
the cranked Nilsson potential for the yrast line and analysis of 
the low-lying excitations near the yrast line in the RPA. 
Our consideration is based on the  Hamiltonian
\begin{equation}
 \hat H_{\Omega} \,= \hat H_0 -\sum_\tau \lambda_{\tau} \hat N_{\tau}
+V-\Omega \hat J_x.
\label{h1}
\end{equation}
The term  $\hat H_0=\hat H_N\,+ \hat H_{\rm add}$ contains the Nilsson Hamiltonian
$\hat H_N$ with quadrupole deformations and the additional term that restores the 
local Galilean invariance of the Nilsson potential, broken in the rotating frame. 
To describe different parity states, the interaction $V$ includes 
separable monopole pairing,  monopole-monopole, quadrupole-quadrupole and  
spin-spin terms for the positive parity $(\pi=+)$  and dipole-dipole, octupole-octupole 
terms for the negative parity $(\pi=-)$. The chemical potentials $\lambda_\tau$ $(\tau=$n or p) 
are determined so as to give correct average particle numbers $\langle \hat N_\tau \rangle$.
All multipole and spin-multipole operators have a good isospin $T$ and signature 
$r=\pm 1$ (see the properties of the matrix elements in Ref.\onlinecite{kv98}). 
They  are expressed in terms of doubly stretched coordinates 
$\tilde x_i = (\omega_i/\omega_0)\,x_i$, which ensure the self-consistent conditions 
at the equilibrium deformation. All the details of this approach are
thoroughly discussed in Ref.\onlinecite{Kv06}. Although the CRPA approach is based on
the Nilsson potential, it contains static and dynamic effects  induced by  rotation 
and serves a useful purpose of independent view of the Skyrme results.

The mode associated with the rotation about the $x$-axis determines the 
Thouless-Valatin moment of inertia
${\cal J_{TV}}$ 
\begin{equation}
\label{tv}
[\hat H_\Omega(^{\pi=+}_{r=+}), i{\hat \Phi}]=
\frac{\hat J_x}{\cal J_{TV}}, \quad [{\hat \Phi},{\hat J}_x]=i.
\end{equation}
Here, $\hat H_\Omega(^{\pi=+}_{r=+})$ is the positive signature term
of the full Hamiltonian; an angle operator ${\hat \Phi}$ is the canonical partner of the
angular momentum operator ${\hat J}_x$.  The solution of these equations leads 
to the definition  ${\cal J_{TV}}=detA/detB$ where the matrix A(B) has a dimension $n=10(11)$
due to a coupling of different operators involved in $\hat H_\Omega(^{\pi=+}_{r=+})$ 
\cite{Kv06}. We recall that a comparison of  the dynamical moment of inertia 
${\cal J}^{(2)}=-d^2E/d\Omega^2$ 
$(E=\langle  \hat H_{\Omega} \rangle$) with the Thouless-Valatin ${\cal J}_{TV}$ 
moment of inertia calculated in the RPA provides a faithful  test of the self-consistency 
of a cranking microscopic theory. The equivalence between these momenta of inertia 
certainly holds, if one found a self-consistent  mean field minimum and spurious 
solutions are separated from the physical ones (see exact results for a solvable 
model in Ref.\onlinecite{n2}). The use of the phenomenological rotational dependence 
of the gap parameter near the backbending region (see Ref.\onlinecite{Kv06}), 
the $ls$ and $l^2$ terms of the Nilsson potential affect this equivalence. 
However the result achieved in our calculations yields a reasonable 
self-consistency (see Fig.3). We obtain a good agreement between our results and 
experimental kinematic 
$\Im^{(1)}_{exp,\,\nu}(\Omega(I))= \hbar^2 2I/(E_{\nu}(I+1) - E_{\nu}(I-1))$ and
the dynamic
$\Im^{(2)}_{exp;\,\nu}=\hbar dI/d\Omega = 2 \hbar/(\Omega_{\nu}(I+1)-\Omega_{\nu}(I-1))$
momenta of inertia for the yrast states. 
Here, $E_{\nu}(I)$ is an experimental energy  for the 
rotational band $\nu$ ($\nu=yrast,\,\beta,\,\gamma,\,...$) \cite{BN}. 
The drastic change of the kinematic momenta of inertia starts slightly earlier 
and ends slightly later than in experiments. It is known that in the backbending 
region the cranking approach is less reliable \cite{ham}. Nevertheless, one observes 
a similar rotational dependence of the magnitude of the calculated and 
experimental kinematic  momenta of inertia till $\sim 0.6$MeV in $^{164}$Yb, 
while a good agreement holds till $\sim 0.45$MeV in $^{162}$Yb (see below).
The results for the dynamic momenta of inertia reproduce with a reasonable accuracy 
the fluctuations in the backbending region, in spite of the approximations made in
our approach.
\begin{figure}[ht]
\includegraphics[height=0.4\textheight,clip]{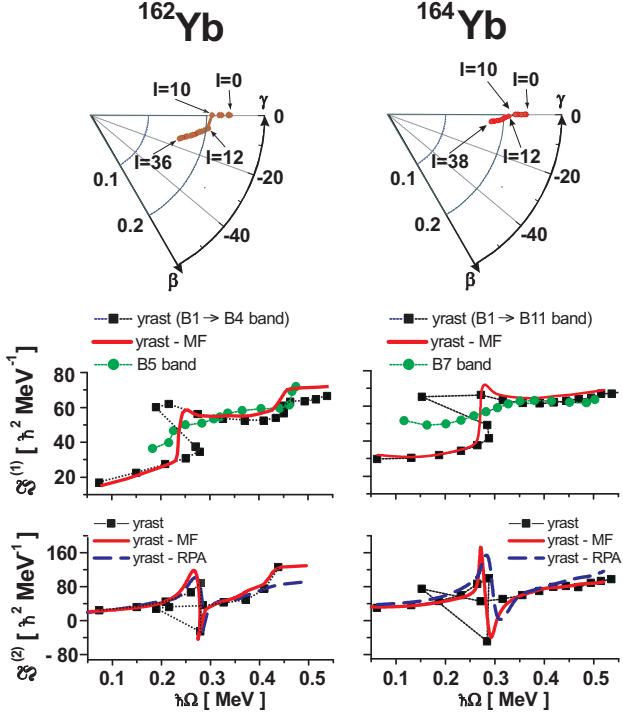}
\caption{(Color online) $^{162}$Yb (left) and $^{164}$Yb (right).
Top panels: equilibrium deformations in $\beta$-$\gamma$ plane
as a function of the angular momentum $I = \langle {\hat J}_x \rangle-1/2$
(in units of $\hbar$) .
Middle panels: the kinematic $\Im^{(1)}(\Omega)=\langle \hat J_x\rangle/\Omega $ 
momenta of inertia. 
Filled square (circle) is used for experimental positive (negative) parity lowest 
states. Different symbols display the experimental data 
associated with B1,B2...bands (the band labels are taken in 
accordance with the definitions given in Ref.\onlinecite{BN}).
Bottom panels: the dynamic $\Im^{(2)}(\Omega)$
momenta of inertia are compared with the corresponding 
experimental values (filled squares). 
Experimental values are connected by thin line to guide eyes.
The calculated values of the kinematic (middle) and dynamic (bottom) momenta of 
inertia for the yrast line are connected by a solid line. The dashed 
line displays the behavior of the Thouless-Valatin moment of inertia (bottom).
All results are obtained within the cranked Nilsson plus RPA (CRPA) approach. 
} 
\label{fig3}
\end{figure}
The consistency between the mean field and the RPA results was achieved by varying 
the strength constants of the pairing and multipole interactions in the 
RPA (see Ref.\onlinecite{Kv06}) to fulfill the conservation laws  for both 
signatures and both parities
\begin{eqnarray}
\left[\hat H_\Omega(^{\pi=+}_{r=+}) \,,\hat N_{\tau} \right]   &=& 0, \,\,\,\,\,\,\,\,
\left[\hat H_\Omega(^{\pi=-}_{r=+}) \,,\hat P_x \right]  = 0, \nonumber\\ 
\left[\hat H_\Omega(^{\pi=+}_{r=+})\,,\,\hat J_x \right] &=& 0,  \,\,\,\,\,\,\,\,
\left[\hat H_\Omega(^{\pi=+}_{r=-}),\hat \Gamma^{\dagger}\right] = \hbar\Omega \hat \Gamma^{\dagger},
\label{7}
\end{eqnarray}
where $\hat \Gamma^{\dagger}=(\hat J_z + i \hat J_y)/\sqrt{2 \langle \hat J_x \rangle}$. 
It results in the separation of physical modes (excitations)  from those that are related 
to the symmetries broken by the mean field.  Two Goldstone modes are associated with 
the violation of the particle number (for protons and for neutrons). The other two 
Goldstone modes are related to the translational and spherical symmetries of the 
mean field. The last equation  yields a negative signature solution $\hbar \Omega$, 
a collective rotational mode arising from the symmetries broken by the external 
rotational field. We recall that the decoupling of  $\hat P_y$ and $\hat P_z$ 
components of the center of mass operator from physical modes is uniquely defined  
in the cranking model with a signature and a parity symmetry \cite{chor}.
The constraint satisfaction of the translational symmetry related to the
operator $\hat P_x$ determines, therefore, the isoscalar dipole strength solely  
and also guarantees the decoupling of the spurious center of mass motion from 
physical modes. This is especially important for the calculations of $B(E1)$-transitions.
For the isovector dipole coupling strength we adopt the standard value 
$\chi_{1}=\pi V_1/A \langle{\tilde r}^2\rangle$ with $V_1=130$MeV \cite{BM75}.
The self-consistent octupole strength constants obtained for the anisotropic harmonic 
oscillator \cite{sk} are used in our calculations for octupole excitations, taking
into account a rotational dependence of different mean field values in the original
expression. By virtue of these constants we nicely reproduce the experimental octupole 
Routhians (see below).

\begin{figure}[ht]
\includegraphics[height=0.4\textheight,clip]{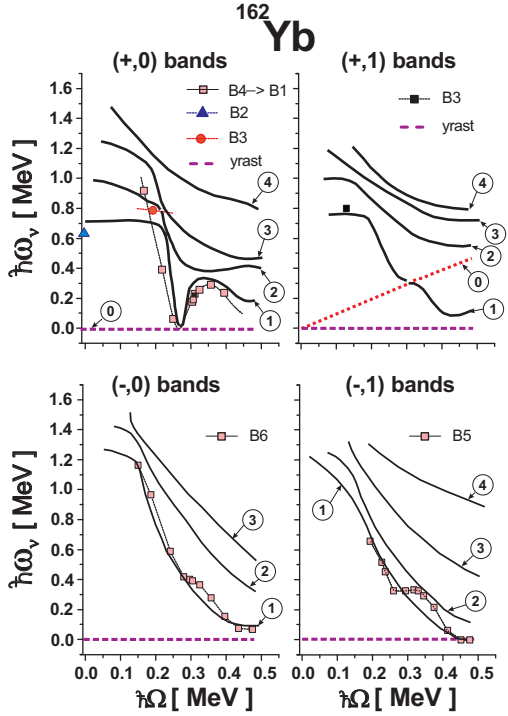}
\caption{(Color online) $^{162}$Yb.
The rotational evolution of CRPA solutions: the positive (negative) signature 
ones with even (odd) spins are displayed on left (right) panels. 
Number in a circle denotes 
the RPA solution number : $1$ is the first $\nu=1$  RPA solution  {\it etc}.
Different symbols display the experimental data 
associated with B1,B2...bands.
The yrast line position denoted as  "0" (top, left) is displayed by
dashed line on all panels.
Top panels: positive  parity  $\pi=+$ solutions. The redundant mode (right)
$\omega_\nu=\Omega$ is denoted as "0" and is displayed by dotted line. 
Bottom panels: negative parity $\pi=-$ solutions. 
} 
\label{fig4}
\end{figure}
The different response on the rotation can be understood from the analysis of excited 
states. To this aim we define the experimental excitation energy in the rotating frame 
$\hbar \omega_{\nu}(\Omega)_{\it exp}=R_{\nu}(\Omega) - R_{yr}(\Omega)$ 
as a function of the rotational frequency $\Omega$ \cite{Na87}. 
Here, the Routhian function 
$R_{\nu}(\Omega) = E_{\nu}(\Omega) - \hbar \Omega\, I_{\nu}(\Omega)$.
The experimental states are classified by the parity $\pi=\pm$ and a quantum number 
$\alpha$ which is equivalent to our signature $r$ (see Figs.4,6). The positive signature states 
($r=+1$) correspond to $\alpha=0$ which characterizes rotational bands with even 
spin in even-even nuclei. The negative signature states ($r=-1$) correspond to 
$\alpha=1$ and are associated with odd spin states in even-even nuclei. The energy 
$\hbar \omega_{\nu}(\Omega)_{\it exp}$ can be compared with the 
RPA results, $\hbar \omega_{\nu}(\Omega)$,  calculated at  a given rotational 
frequency. The agreement between our results and experimental data indicates that 
quadrupole deformation (axial and nonaxial) is a major factor that determines the 
evolution of the  mean field till $\hbar\Omega\approx 0.6(0.45)$MeV 
in $^{164}$Yb ($^{162}$Yb). 
\begin{figure}[ht]
\includegraphics[height=0.29\textheight,clip]{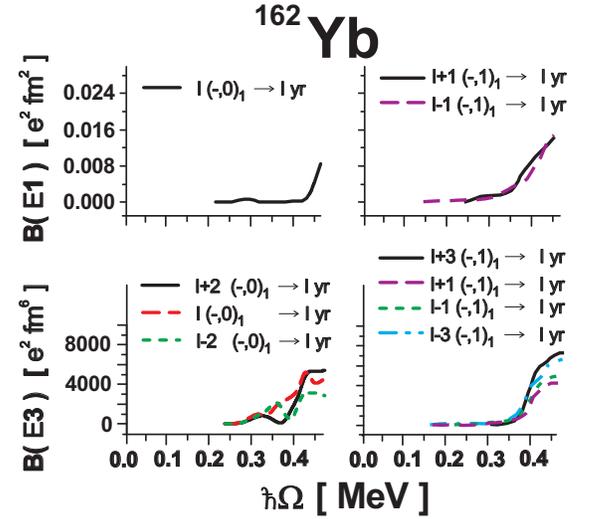}
\caption{(Color online) $^{162}$Yb. 
Reduced $B(E1)$- and $B(E3)$- transition probabilities 
from the lowest negative parity one-phonon  states to the yrast states. 
We used the effective proton (neutron) charge $e=N/A(-Z/A)$ for the calculation
of $B(E1)$-transitions, while only the proton contribution is taken into account
for the $B(E3)$- transitions.
Left(right) panels correspond to the transition with  $\Delta I$ even 
($\Delta I$ odd).
All results are obtained within the CRPA approach. 
} 
\label{fig5}
\end{figure}

In  $^{162}$Yb the band B4 crosses the band B1 at the rotational frequency 
$\hbar \Omega = 0.265$MeV (a transition point) and becomes the yrast one for 
$\hbar \Omega > \, 0.265$MeV. The change of the yrast line structure produces 
the backbending (see  Fig.3). It should be emphasized that the potential-energy surfaces 
are very shallow. The deformations shown in Fig.3 should be not taken as precise 
predictions, since even before the transition point the yrast states may possess 
small $(\sim 1^0-3^0)$ $\gamma$-deformation (see also Refs.\onlinecite{fr90,emr}).
To elucidate  the structure near the transition 
point we trace the rotational evolution of the CRPA solutions and the lowest 
two-quasiparticle poles. At $\Omega=0$ each quasiparticle orbital (routhian) is 
characterized by asymptotic Nilsson quantum numbers. These numbers are irrelevant, 
however, in the rotating case and are used only for convenience. The first positive 
parity and signature RPA solution (see Fig.4, top left panel) may be identified with 
$\beta$-excitations at $\hbar \Omega\leq 0.2$MeV. The increase of the rotational 
frequency leads  to a strong mixing between $\gamma$- (the second RPA solution at low spins) 
and $\beta$-excitations. At the transition point from an axial to nonaxial rotations,
$\hbar\Omega\approx 0.26$MeV, a two-quasiparticle neutron component $3/2[651]3/2[651]$
dominates in the quadrupole phonon structure ($\sim 92\%$). This fact suggests 
that the alignment of a pair $i_{13/2}$ is the main mechanism that
drives the nucleus to triaxial shapes. The first negative signature RPA solution 
(Fig.4, top right panel) carries the properties of a  gamma-vibrational mode (with 
odd spins) till $\hbar\Omega\approx 0.28$MeV and does not lead to any instability.
\begin{figure}[ht]
\includegraphics[height=0.4\textheight,clip]{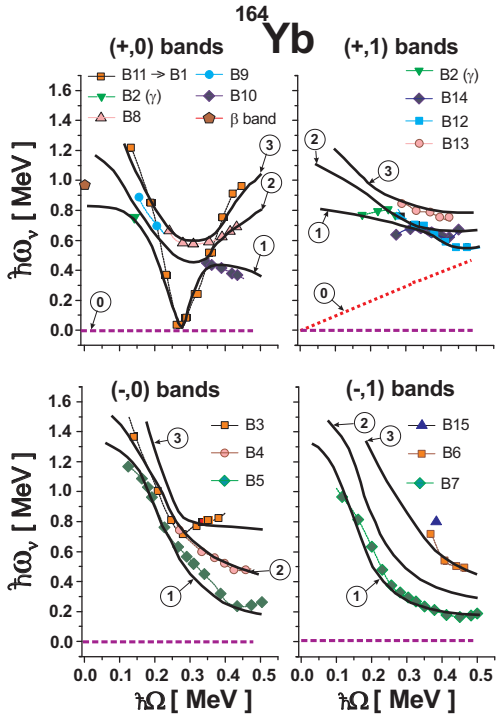}
\caption{(Color online) $^{164}$Yb.
Similar to Fig.4.
} 
\label{fig6}
\end{figure}

From Fig.4 (bottom panel, right) one observes that the first negative signature 
and parity RPA (octupole) solution tends to zero.  Almost a zero energy gap between 
the  negative parity band (B5) and the yrast line is also observed at 
$\hbar \Omega \geq 0.45$MeV. The collectivity of the lowest  negative parity RPA 
solutions of both signatures increases noticeably with the increase of the rotational 
frequency. Results for reduced $B(E1)$- and $B(E3)$-transition probabilities evidently 
demonstrate the onset of octupole correlations  at  $\hbar \Omega > 0.3$MeV 
(see Fig.5). For example, $B(E3,\Delta I=3)$-transition probability increases from
$\sim 0.7$ W.u. at $\hbar\Omega\approx 0.3$ MeV to $\sim 4.5$ W.u. 
at $\hbar\Omega\approx 0.45$ MeV (here W.u is the Weisskopf unit).
Several two-quasiparticle components  originated from $h_{11/2}$ and 
$g_{7/2}$ subshells for protons and $i_{13/2}$ and $h_{9/2}$ subshells for neutrons 
contribute to the collectivity of the lowest negative parity one-phonon states. 
The maximal  weight of two-quasiparticle components is  $\sim 65\%$. All these 
features reveal the nature of a shape transition at $\hbar \Omega \sim 0.45$MeV. 
The onset of the static octupole deformations becomes feasible, since the instability 
point found in the CRPA approach coincides with the result of the Skyrme mean field 
calculations (see Fig.1). These results suggest that the octupole deformations
are due to  the octupole phonon condensation at fast rotation.
\begin{figure}[ht]
\includegraphics[height=0.29\textheight,clip]{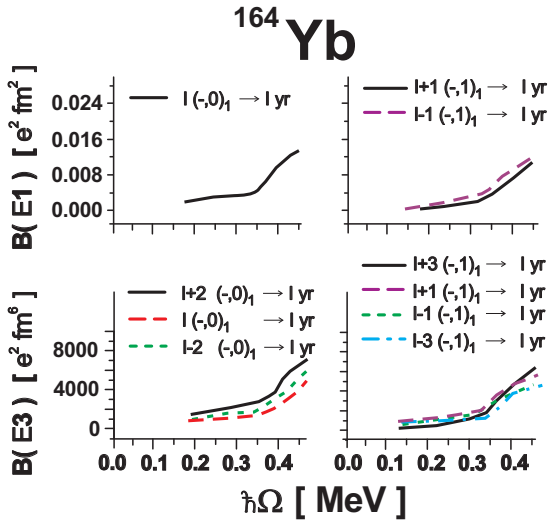}
\caption{(Color online) $^{164}$Yb. 
Similar to Fig.5.
} 
\label{fig7}
\end{figure}

In $^{164}$Yb (see Fig.6), there are observed 15 rotational bands so far \cite{BN}.
The positive signature family (with even spin) is comprised of B1, B2, B8, B9, B10, 
B11 bands of positive and B3, B4, B5 bands of negative parities. The negative 
signature family (with odd spins) includes B2, B12, B13, B14, and B6, B7, B15 of 
positive and negative parity bands, respectively. At $\hbar \Omega \approx 0.27$MeV 
the band B11 crosses the ground band B1. This crossing gives rise to the backbending 
(see Fig.3). According to our analysis, the first positive parity and signature RPA solution 
(see Fig.6, top left panel) may be identified with $\gamma$-excitations at 
small rotation $\hbar \Omega\leq 0.2$MeV. The rotation induces a strong mixing 
between $\gamma$-and $\beta$-excitations (the second RPA solution at low spins). 
Although the pattern is similar to the one of $^{162}$Yb,  $\beta$- and $\gamma-$
vibrations are interchanged at small rotation. At $\hbar\Omega\approx 0.27$MeV 
the transition from an axial to nonaxial rotations is determined by two-quasiparticle
neutron component from $i_{13/2}$ subshell. A good agreement between data and our 
results implies that quadrupole degrees of freedom dominate in this domain of  
rotational frequency values. The negative parity RPA solutions of both signatures 
(see Fig.6) decrease with the rotational frequency. The RPA results reproduce with 
a good accuracy the rotational behaviour of the negative parity states, especially,
the lowest one. While the octupole phonons are of collective nature 
(the maximal component is $\sim 75\%$), the quadrupole deformed field remains stable, 
even at very high spins. The growing octupole collectivity  is exhibited in the 
increasing dipole and octupole transitions from the negative-parity one phonon
states at $\hbar\Omega > 0.3$MeV (see Fig.7). 
In particular, $B(E3,\Delta I=3)$-transition probability increases from
$\sim 1.2$ W.u. at $\hbar\Omega\approx 0.3$ MeV to $\sim 4.4$ W.u. 
at $\hbar\Omega\approx 0.45$ MeV. In contrast with $^{162}$Yb, all 
negative parity solutions are nonzero at $\hbar\Omega\leq0.6$MeV. Octupole correlations 
are unable to break the reflection symmetry in $^{164}$Yb, which elucidates the 
result of Skyrme calculations (see Fig.1). 

Summarizing, we found that the internal structure of the low-lying excited 
($\pi=+$, $r=+1$) band, which crosses the ground band, has almost a pure 
two-quasiparticle character  originating from $i_{13/2}$ neutron subshell, in 
agreement with the prediction of the phenomenological analysis \cite{fr90}. 
In both nuclei the rotation produces a profound affect on  shell structure inducing 
noticeable octupole correlations at high spins. 
At $\hbar\Omega > 0.3$MeV we predict the formation of  
octupole bands with strong $B(E1)$- and $B(E3)$- transitions to the yrast line  
(see Figs.5,7). We found that at $\hbar \Omega \approx 0.45$MeV the  
octupole phonon solution vanishes in the rotating frame in $^{162}$Yb. It results 
in  the onset of the degeneracy between the lowest negative parity and negative 
signature band and the positive parity and positive signature yrast line. 
This mechanism explains the  spontaneous 
breaking of reflection symmetry of the rotating mean field with Skyrme interaction 
(see Fig.1). In contrast, in $^{164}$Yb the octupole correlations manifest themselves 
as a low-lying octupole vibrations of the quadrupole deformed rotating nucleus, 
in agreement with the Skyrme results.

\section*{Acknowledgments}
This work is a part of the research plan MSM
0021620859 supported by the Ministry of Education of the Czech Republic
and by the project 202/06/0363 of Czech Grant Agency.
It is also partly supported by Grant No. FIS2005-02796 (MEC, Spain).
R. G. N. gratefully acknowledges support from the
Ram\'on y Cajal programme (Spain).

\end{document}